\documentclass[aps,showpacs,twocolumn]{revtex4-1}
\usepackage[T1]{fontenc}
\usepackage{amssymb,amsfonts,amsmath,mathtext,enumerate,float,mathrsfs}
\usepackage[utf8]{inputenc}
\usepackage[english]{babel}
\usepackage{graphicx,color}
\usepackage{wrapfig}
\graphicspath{{./}}

\usepackage{epsfig}
\usepackage{epsf}
\usepackage{subfig} 
\usepackage{feynmp} 
\usepackage{pgf,pgfarrows,pgfnodes,pgfautomata,pgfheaps,multirow}
\usepackage{array}
\usepackage{bm}
\usepackage{latexsym}
\usepackage{pifont}

\newcommand{\be}{\begin{equation}}
\newcommand{\ee}{\end{equation}}
\newcommand{\bea}{\begin{eqnarray}}
\newcommand{\eea}{\end{eqnarray}}
\newcommand\nn{\nonumber}

\begin{document}


\title{Fixing the vector coupling constant $g_{\rho} = 5.0$ in the NJL model and
final-state interactions in processes $\rho \to e^{+}e^{-}[\mu^{+}\mu^{-}]$, 
$\rho \to \pi^{+}\pi^{-}$, $\tau \to \pi^{-}\pi^{0}\nu_{\tau}$, 
and $e^{+}e^{-}\to\pi^{+}\pi^{-}$}

\author{M.\ K.\ Volkov}
\email{volkov@theor.jinr.ru}
\affiliation{Bogoliubov Laboratory of Theoretical Physics, JINR, Dubna, 141980 Russia}

\author{A.\ B.\ Arbuzov}
\email{arbuzov@theor.jinr.ru}
\affiliation{Bogoliubov Laboratory of Theoretical Physics, JINR, Dubna, 141980 Russia}

\author{K.\ Nurlan}
\email{nurlan.qanat@mail.ru}
\affiliation{Bogoliubov Laboratory of Theoretical Physics, JINR, Dubna, 141980 Russia}
\affiliation{Al-Farabi Kazakh National University, Almaty, 050040, Republic of Kazakhstan}
\affiliation{Eurasian National University, Nur-Sultan, 01008, Republic of Kazakhstan}

\author{A.\ A.\ Pivovarov}
\email{tex$\_$k@mail.ru}
\affiliation{Bogoliubov Laboratory of Theoretical Physics, JINR, Dubna, 141980 Russia}

\begin{abstract}
The possibility to use the width of the decay $\rho \to e^{+}e^{-}$
to fix the input parameter $g_\rho=5.0$ of the $SU(2) \times SU(2)$ chiral-symmetric 
Nambu--Jona-Lasinio model is discussed. It is shown that for 
a consistent simultaneous description of the processes 
$\rho \to e^{+}e^{-}$, $\rho \to \pi^{+}\pi^{-}$, $\tau^{-} \to \pi^{-}\pi^{0} \nu_{\tau}$, 
and $e^{+}e^{-} \to \pi^{+}\pi^{-}$ can be constructed.
Taking into account the interaction of pions in the final state appears to be important. 
The obtained theoretical results for the considered processes 
are in a satisfactory agreement with experimental data.
\end{abstract}

\date{\today}



\maketitle

\section{Introduction}

In the version of the Nambu--Jona-Lasinio (NJL) model formulated in~\cite{Volkov:1986zb}, 
the experimental values of the decay widths $\pi^{\pm} \to \mu^{\pm} \nu$ ($F_{\pi}=92.4$ MeV) 
and $\rho \to \pi^{+}\pi^{-}$ ($g_{\rho}=6.14$) were used to fix such important parameters 
as the constituent masses of light quarks and the ultraviolet cutoff parameter. 
These values were obtained in the leading order approximation in the
$1/N_c$ expansion where $N_c$ is the number of colors in QCD. 
A similar approximation was used in the construction of some other versions of the NJL model 
\cite{Ebert:1985kz,Vogl:1991qt,Klevansky:1992qe,Hatsuda:1994pi}. Within the framework 
of these models, it was possible to describe many internal properties of mesons as well as 
the main types of strong, electromagnetic, and weak processes of interaction of mesons at 
low energies. However, for a number of very important processes such as 
$e^{+}e^{-} \to \pi^{+}\pi^{-}$ and $\tau^{-} \to \pi^{-}\pi^{0} \nu_{\tau}$, 
it was not possible to get a satisfactory agreement with experimental data 
using the NJL model within the indicated approximation. 
We assume that this is a consequence of the fact that  
the interaction of pions in the final state in these processe plays an important
role. These interactions can be described by taking into 
account the exchange of outgoing pions by the $\rho$ meson (P-wave meson loop)
as was demonstrated in our recent work~\cite{Volkov:2020uld}. 
That required going beyond the $1/N_c$ expansion of the standard NJL model.

In the present paper we consider the possibility to define the $g_\rho$
coupling constant from the experimental data on the decay width
$\rho \to e^{+}e^{-}$ instead of $\rho \to \pi^{+}\pi^{-}$.
As we will see, the redefinition of this coupling constant leads
to shifts in other important parameters of the NJL model.

\section{Interaction Lagrangian of the NJL model}
\label{Sect:2}

The Lagrangian of the NJL model, describing the interactions of $\pi$, $\rho$, 
and $\omega$ mesons with quarks, has the form~\cite{Volkov:1986zb}
\begin{eqnarray}
\label{L1}
&& \Delta{\mathcal L}_{int} = \bar{q}\biggl[ ig_{\pi}\gamma_5\tau_3\pi^{0} 
+ \frac{g_\rho}{2}\left(I \gamma_{\mu}{\omega}_{\mu} + \tau_3\gamma_{\mu}{\rho^0}_{\mu} \right) 
\nn \\ && \qquad
+ \frac{g_\rho}{2}\gamma_{\mu}\tau_{-}{\rho^-}_{\mu} + ig_{\pi}\gamma_5\tau_{-}\pi^{-}  \biggr]q,
\end{eqnarray}
where $q$ and $\bar{q}$ are fields of $u$ and $d$ quarks;
${\rho}, {\omega}, \pi$ are meson fields; $\tau_3$ is the Pauli matrix,
$\tau_{-}=(\tau_1+i\tau_2)/\sqrt{2}$ is the linear 
combination of the Pauli matrices; $I$ is the identity matrix. 
The values of the quark-meson interaction constants are fixed by the experimental 
decay widths $\pi^{\pm} \to \mu^{\pm}\nu$ ($F_{\pi}=92.4$~MeV) and 
$\rho \to e^{+}e^{-}$ ($g_{\rho}=5.0$) 
(see Sect.~\ref{Sect:3}):
\begin{eqnarray}
\label{Couplings}
g_{\rho} = \sqrt{\frac{3}{2I_{2}}} \approx 5.0, 
&\quad& 
g_{\pi} = \frac{m_u}{F_{\pi}}=\sqrt{\frac{Z_{\pi}}{6}}g_{\rho},
\end{eqnarray}
where
\begin{eqnarray}
Z_{\pi} = \left(1 - 6\frac{m^2_u}{M^{2}_{a_{1}}}\right)^{-1},
\end{eqnarray}
here $Z_{\pi}$ is additional renormalization constants appearing in the transitions 
between axial-vector and pseudoscalar mesons \cite{Volkov:1986zb}, 
$M_{a_{1}} = 1230$~MeV is the mass of the $a_{1}(1260)$ meson~\cite{Tanabashi:2018oca}.
	
Integrals appearing in quark loops are
\begin{eqnarray}
I_{2} = -i\frac{N_c}{(2\pi)^{4}}\int\frac{\Theta(\Lambda_{4}^{2} + k^2)}{(m^{2} - k^2)^{2}}
\mathrm{d}^{4}k  \\  \nonumber
= \frac{N_c}{(4\pi)^2}\left[ \ln\left(1+ \frac{\Lambda^2_{4}}{m^2} \right) 
- \frac{\Lambda^2_{4}}{\Lambda^2_{4}+m^2} \right],
\end{eqnarray}
where $\Lambda_{4}$ is the cutoff parameter \cite{Volkov:1986zb}, 
$N_c = 3$ is the number of colors in QCD. 
	
Solving the reduced equations (\ref{Couplings}) for the mass $m_{u}$ and the cutoff parameter 
$\Lambda_4$, we get:
\begin{eqnarray}
m^2_{u} = \frac{{M}^2_{a_1}}{12}\left[ 1- \sqrt{1- {\left({\frac{2 g_{\rho} F_{\pi}}{M_{a_1}}}\right)}^2} \right] 
\nn \\ 
\Rightarrow  m_{u} = 207~\mathrm{MeV}\ \ \mathrm{and}\ \ \Lambda_4 = 1630~\mathrm{MeV}.
\end{eqnarray}

\section{Decays $\rho \to e^{+}e^{-}[\mu^{+}\mu^{-}]$ and $\rho \to \pi^{+}\pi^{-}$}
\label{Sect:3}

\begin{figure}[h]
\center{\includegraphics[scale = 0.2]{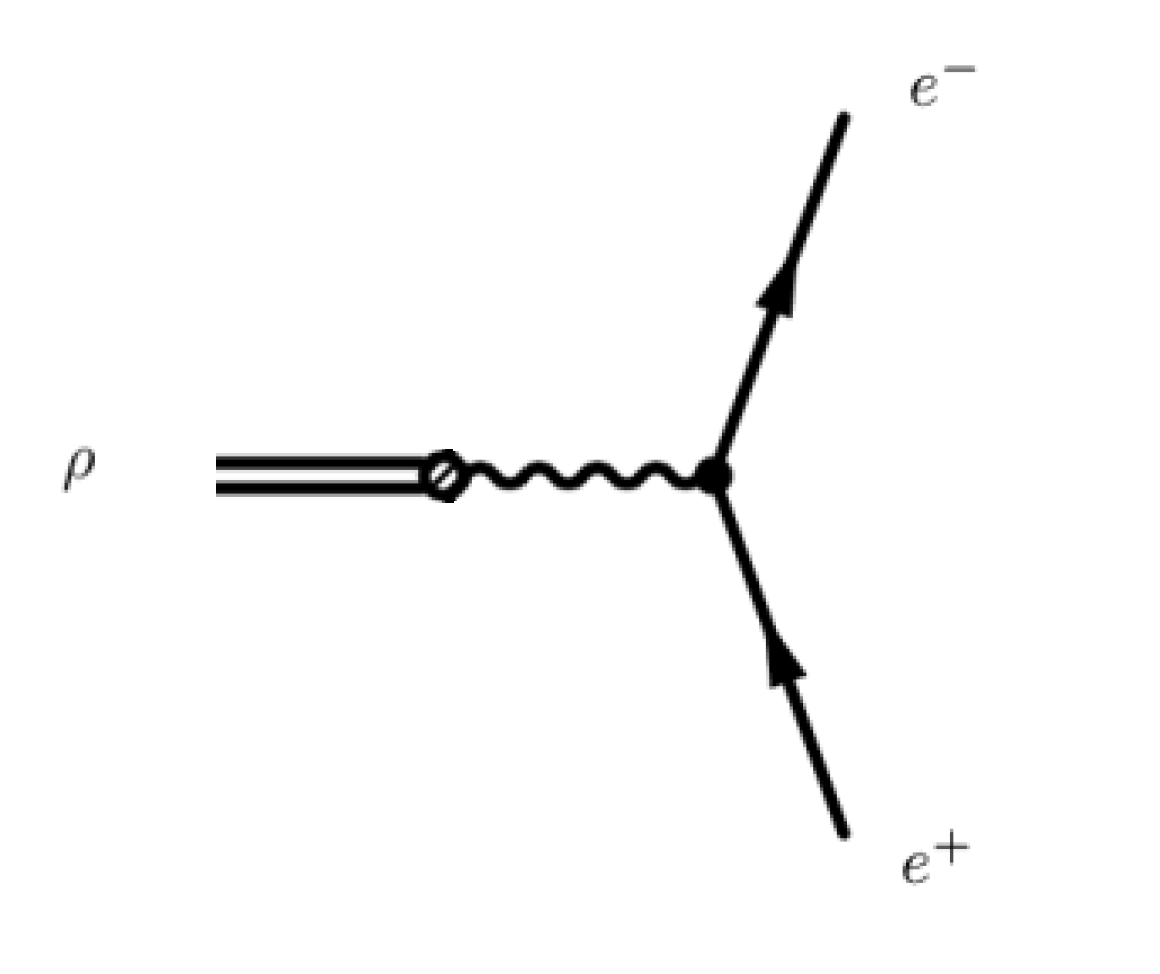}}
\caption{Diagram of the decay $\rho \to e^{+}e^{-}$.}
\label{diagram1}
\end{figure}

In the NJL model the electromagnetic decay $\rho \to e^{+}e^{-}$ is described by the diagram
shown in Fig.~\ref{diagram1}. The corresponding amplitude reads	
\begin{eqnarray}
\mathcal{M} (\rho \to e^{+}e^{-})= \frac{4\pi \alpha_{em}}{g_{\rho}}l_{\mu}e_{\mu}.
\end{eqnarray}
here $\alpha_{em} = 1/137$, $l_{\mu} = \bar{e}\gamma_{\mu} e$ is electromagnetic lepton current, 
$e_{\mu}$ is the polarization vector of the $\rho$ meson.
	
The interaction constant $g_{\rho}$ is determined from the experimental value of the decay 
$\Gamma_{exp}(\rho \to e^{+}e^{-}) = (6.97 \pm 0.07)$ keV \cite{Tanabashi:2018oca}:
\begin{eqnarray}
\Gamma(\rho \to e^{+}e^{-}) = M_{\rho} \frac{4 \pi \alpha^2_{em}}{3 g^2_{\rho}}  
\  \Rightarrow \  g_{\rho}=5.0.
\end{eqnarray}
	
Similar calculations can be done for the $\rho \to {\mu}^{+}{\mu}^{-}$ decay. There is also
a satisfactory agreement with experimental value
$\Gamma_{exp}(\rho \to {\mu}^{+}{\mu}^{-}) = (6.78 \pm 0.42)$~keV 
\cite{Tanabashi:2018oca} for the chosen parameter $g_{\rho} = 5.0$.
	
\begin{figure}[h]
\center{\includegraphics[scale = 0.15]{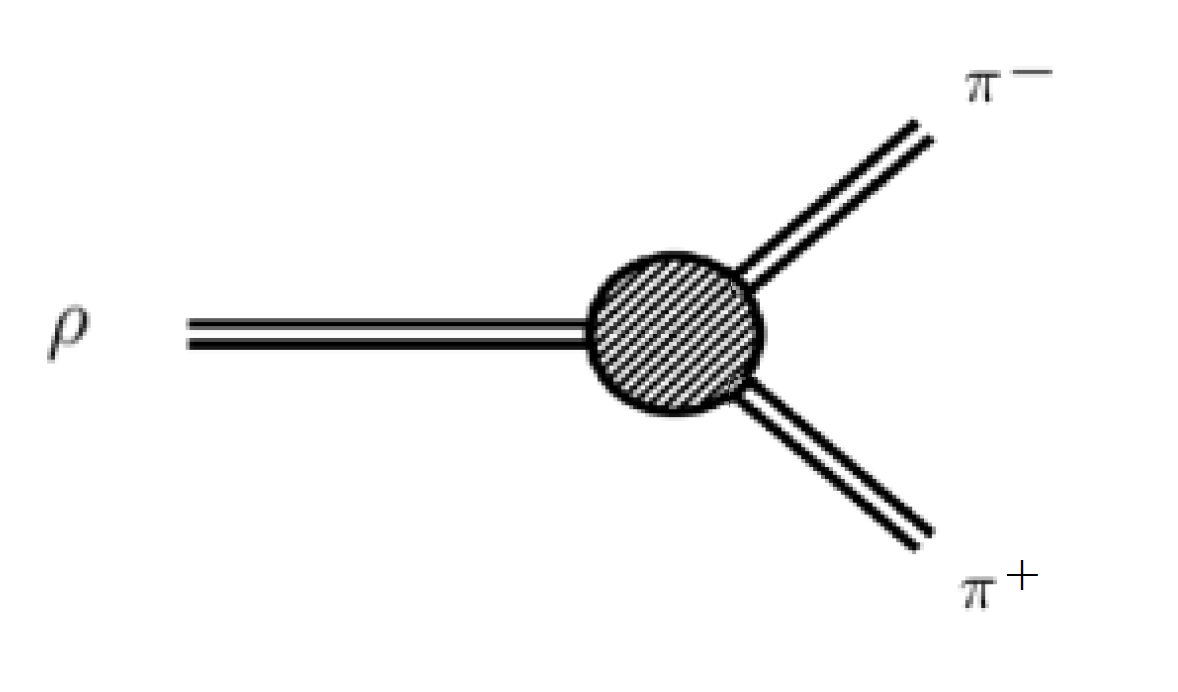}}
\caption{Diagram of the decay $\rho \to \pi^{+}\pi^{-}$.}
\label{diagram2}
\end{figure}

\begin{figure}[h]
\center{\includegraphics[scale = 0.1]{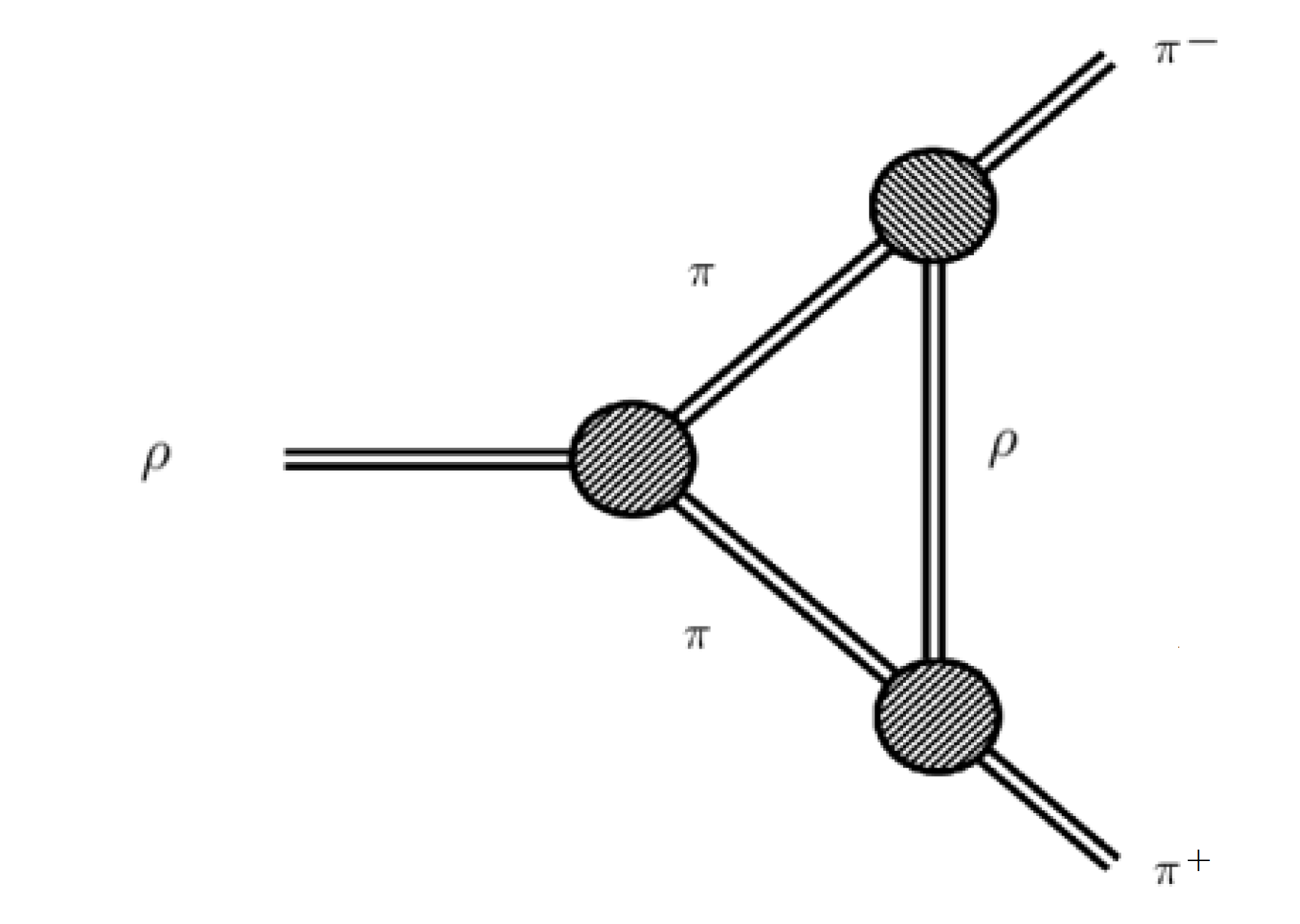}}
\caption{The final state interactions of pions in case of decay $\rho \to \pi^{+}\pi^{-}$.}
\label{diagram3}
\end{figure}

The diagrams of the process $\rho \to \pi^{+}\pi^{-}$ are shown in Figs.~\ref{diagram2}
and \ref{diagram3}.
The amplitude of the decay has the form \cite{Volkov:1986zb}:
\begin{eqnarray}
\mathcal{M}(\rho \to \pi^{+}\pi^{-}) = g_{\rho} e_{\mu} (p_{+} - p_{-})_{\mu}.
\end{eqnarray}	
where $p_{+}$, $p_{-}$ are the pion momentum.  
	
The experimental value for the decay width is $\Gamma(\rho \to \pi^{+}\pi^{-}) = 147.8 \pm 0.9$ MeV 
\cite{Tanabashi:2018oca}. If we disregard interactions in the final state, for the width we get the value 
$\Gamma(\rho \to \pi^{+}\pi^{-}) = 100$~MeV. In order to obtain an agreement with the experiment, it is 
necessary to take into account the pion interactions in the final state. This can be 
carried out by considering the meson loop, shown in Fig.~\ref{diagram3}. Within the NJL model,
this method was introduced in our recent work~\cite{Volkov:2020uld}.
As a result, we get the amplitude 
\begin{eqnarray}
\label{rho}
\mathcal{M}(\rho \to \pi^{+}\pi^{-}) = g_{\rho} \left[1+ T_{\rho\pi\pi} \right] 
 e_{\mu} (p_{+} - p_{-})_{\mu},
\end{eqnarray}	
where 
\begin{eqnarray}
T_{\rho\pi\pi} =  g^2_{\rho} \left[ \frac{I_{1M}}{M^2_{\rho}} + I_{2M} \right].
\end{eqnarray}
		
The second term in amplitude (\ref{rho}) in square brackets corresponds to the interaction of pions 
in the final state, $I_{1m}$ and $I_{2m}$ are quadratically and logarithmically 
divergent integrals,
\begin{eqnarray}
\label{integral}
&& I_{2M} = -i\frac{N_{c}}{(2\pi)^{4}}\int\frac{\Theta(\Lambda_{M}^{2} + k^2)}
{(M^2_{\rho} - k^2)({M^2_{\pi}} - k^2)}\mathrm{d}^{4}k =  \frac{1}{(4\pi)^2} 
\nn \\ \nonumber
&\times& \frac{1}{M^2_{\rho} - M^2_{\pi}} \left[M^2_{\rho} \ln\left(1+ \frac{\Lambda^2_{M}}{M^2_{\rho}} \right) 
- M^2_{\pi} \ln\left(1+ \frac{\Lambda^2_{M}}{M^2_{\pi}} \right) \right],
\end{eqnarray}
\begin{eqnarray}
&& I_{1M} = -i\frac{N_{c}}{(2\pi)^{4}}\int\frac{\Theta(\Lambda_{M}^{2} + k^2)}{(M^2_{\rho} - k^2)}
\mathrm{d}^{4}k =  \frac{1}{(4\pi)^2}  
\nn \\ \nonumber
&\times& \left[ \Lambda^2_{4} - M^2_{\rho} \ln\left(1+ \frac{\Lambda^2_{M}}{M^2_{\rho}} \right) \right],
\end{eqnarray}

The cutoff parameter $\Lambda_M$ in the meson loop integrals can be fixed  
to match the experimental value $\Gamma_{exp}(\rho \to \pi^{+}\pi^{-}) = 147.8 \pm 0.9$~MeV 
\cite{Tanabashi:2018oca}. This gives $\Lambda_M = 880$~MeV.

\section{Processes $\tau^{-} \to \pi^{-}\pi^{0} \nu_{\tau}$ and $e^{+}e^{-} \to \pi^{+}\pi^{-}$ in the NJL model}	
   
In this section we will show that it is possible to get theoretical predictions for the decay width 
$\tau^{-} \to \pi^{-}\pi^{0} \nu_{\tau}$ and the section $e^{+}e^{-} \to \pi^{+}\pi^{-}$ using the parameters 
of our model obtained in the previous Sect.~\ref{Sect:2} and~\ref{Sect:3}. However, if in the previous
work~\cite{Volkov:2020uld} the parameter $\Lambda_{M} = 740$~MeV was fixed according to an experiment
related to measuring the cross section of the process $e^{+}e^{-} \to \pi^{+}\pi^{-}$, here the obtained
theoretical results describing these processes can be considered as predictions.
	
\begin{figure}[h]
\center{\includegraphics[scale = 0.25]{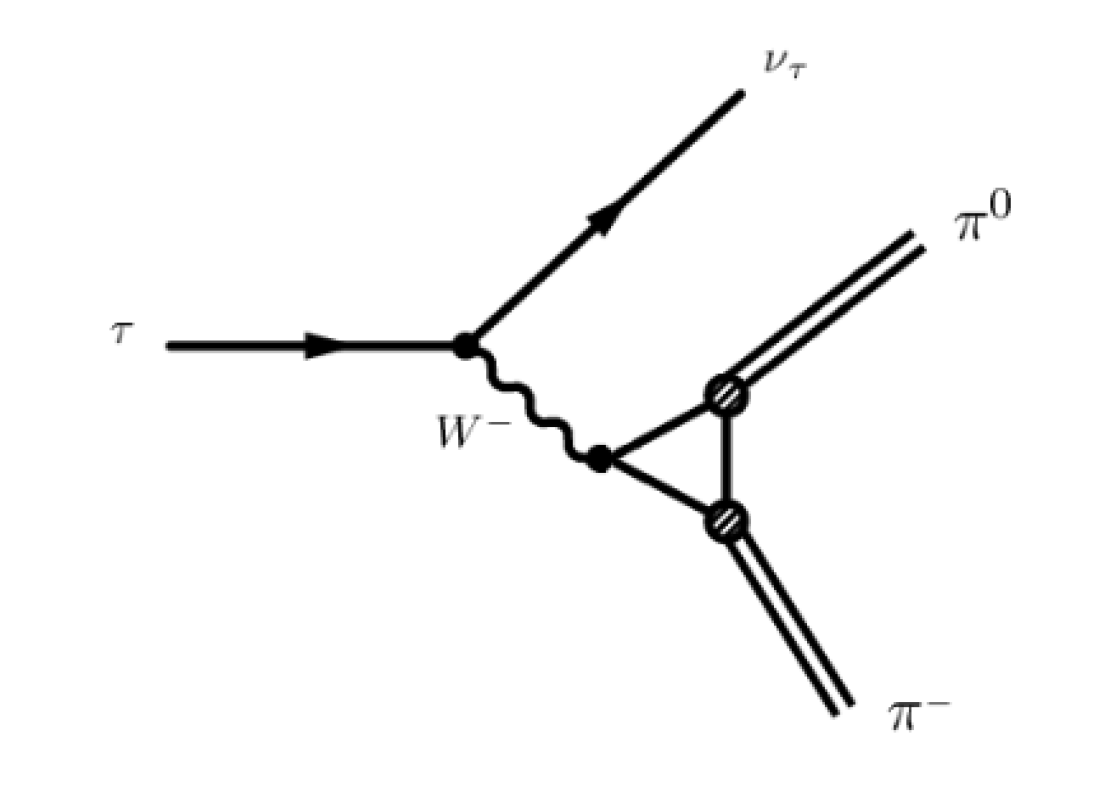}}
\caption{Diagram of the decay $\tau^{-} \to \pi^{-}\pi^{0} \nu_{\tau}$.}
\label{diagram4}
\end{figure}

\begin{figure}[h]
\center{\includegraphics[scale = 0.25]{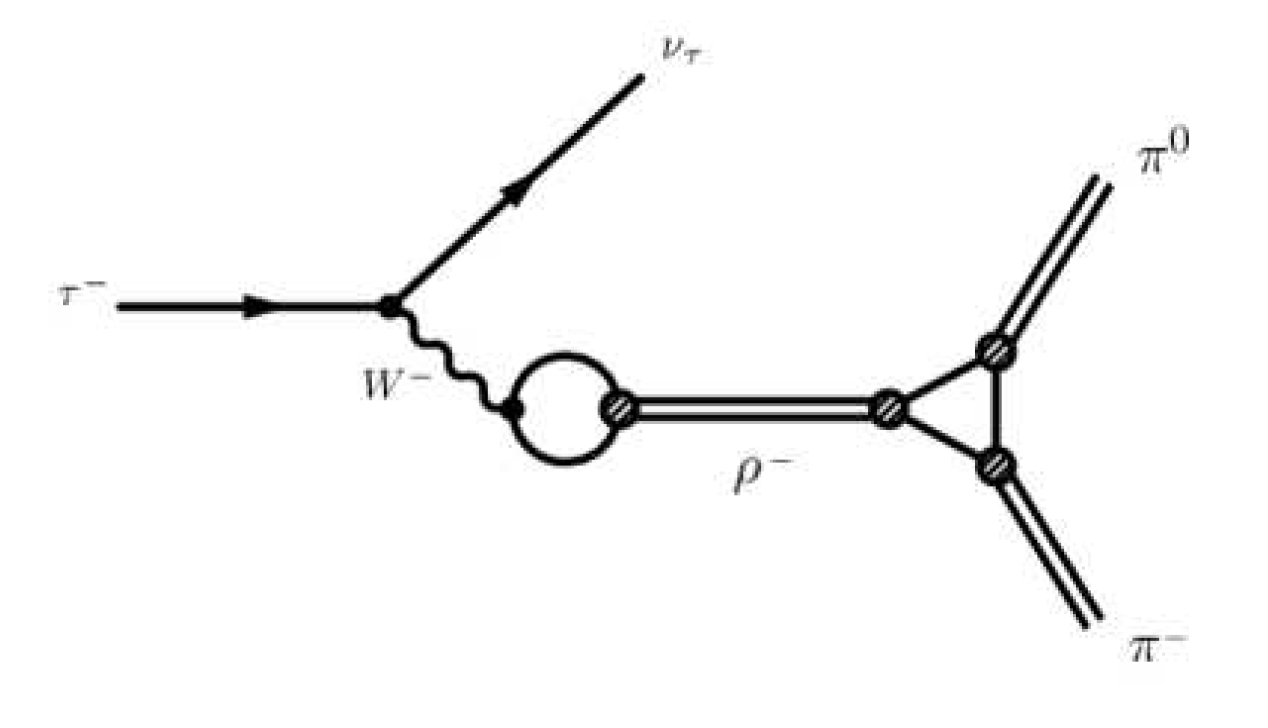}}
\caption{Diagram of the decay $\tau^{-} \to \pi^{-}\pi^{0} \nu_{\tau}$ with intermediate $\rho$ meson.}
\label{diagram5}
\end{figure}

The amplitude of the decay $\tau^{-} \to \pi^{-}\pi^{0} \nu_{\tau}$ in the NJL model takes the 
form (see Figs.~\ref{diagram4} and~\ref{diagram5}):
\begin{eqnarray}
&& \mathcal{M}(\tau^{-} \to \pi^{-}\pi^{0}\nu_{\tau}) =
- G_F V_{ud} \left(1+ T_{\rho} \right)   
\nn \\ \nonumber
&\times&  \left[1+ T_{\rho\pi\pi} \right] \cdot L^{weak}_{\mu} (p_{+} - p_{-})_{\mu},
\end{eqnarray}
where $G_{F}$ is the Fermi constant; $V_{us}$ is the element of the Cabibbo-Kobayashi-Maskawa matrix;
$L^{weak}_{\mu} = \bar{\nu_{\tau}}\gamma_{\mu}(1-\gamma_5)\tau^-$ is the weak lepton current; 
$T_{\rho}$ is the contributions from the intermediate vector $\rho$ meson. 

As a result, for the branching fraction of the decay $\tau^{-} \to \pi^{-}\pi^{0} \nu_{\tau}$ we obtain 
\begin{eqnarray}
\mathrm{Br}_{\mathrm{NJL}}(\tau^{-} \to \pi^{-}\pi^{0} \nu_{\tau}) = (24.7 \pm 1.3) \%. 
\end{eqnarray}
The experimental value for the branching fraction of this decay~\cite{Tanabashi:2018oca} is
\begin{eqnarray}
\mathrm{Br}_{\mathrm{exp}}(\tau^{-} \to \pi^{-}\pi^{0} \nu_{\tau}) = (25.49 \pm 0.09 )\%. 
\end{eqnarray}
Note that the accuracy of the $SU(2) \times SU(2)$ NJL model is about 5\%
which follow from the non-conservation of the axial current (PCAC principle) \cite{Vainshtein:1970zm}
and the comparison of the model predictions with experimental data on low-energy
meson interactions~\cite{Volkov:1986zb}.

\begin{figure}[h]
\center{\includegraphics[scale = 0.45]{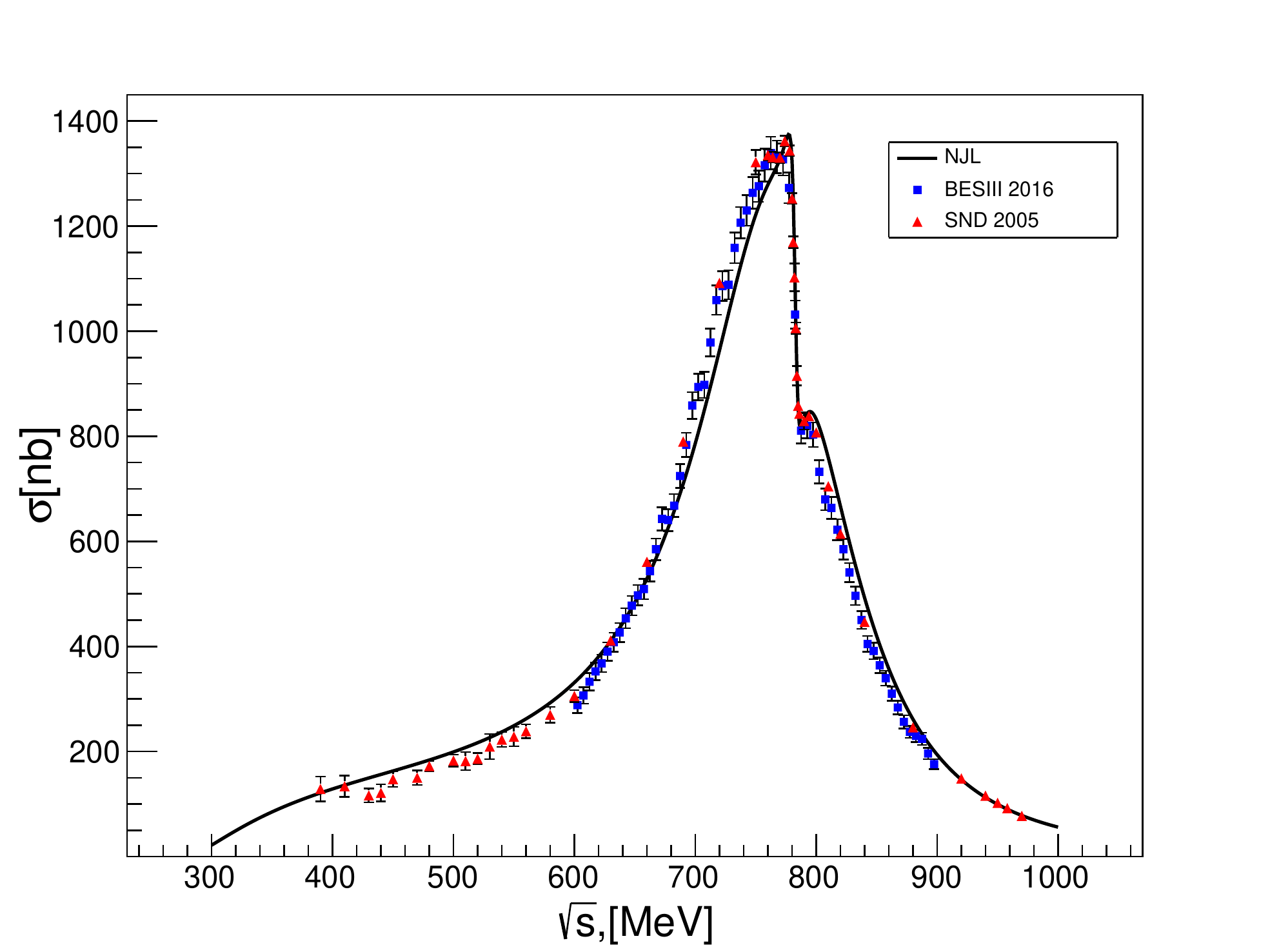}}
\caption{Cross section of the process $e^{+}e^{-} \to \pi^{+}\pi^{-}$ depending on the energy 
from the center-of-mass system. The experimental points are taken from 
\cite{Achasov:2005rg, Ablikim:2015orh}}.
\label{fig:6}
\end{figure}

The amplitude of the process $e^{+}e^{-} \to \pi^{+}\pi^{-}$  in the NJL model with taking into account 
the interaction of pions in the final state reads
\begin{eqnarray}
&& \mathcal{M}(e^{+}e^{-} \to \pi^{+}\pi^{-}) =
- \frac{4 \pi \alpha_{em}}{s} \left(1 + T_{\rho} + T_{\omega} \right) 
\nn \\ \nonumber
&\times& \left[1+ T_{\rho\pi\pi} \right] \cdot L^{em}_{\mu} (p_{+} - p_{-})_{\mu}.
\end{eqnarray}
To describe this process, it is necessary to take into account the mixing of $\omega$ and 
$\rho$ mesons. Within the NJL model the mixing is proportional to the difference of light quark masses.
We used $m_d - m_u = 4$~MeV obtained in the NJL model from the analysis of 
the decay $\omega \to \pi^{+}\pi^{-}$ \cite{Volkov:1986zb}. The corresponding values 
of contributions from intermediate mesons $\rho$ and $\omega$ are taken from \cite{Volkov:2020uld}.
The results obtained for the process  $e^{+}e^{-} \to \pi^{+}\pi^{-}$ when using the new values of
the NJL model parameters are shown in Fig.~\ref{fig:6}.

\section{Conclusions}

In a number of previous works, two alternative possible values for the constant $g_{\rho}$
were considered. 
Namely, $g_{\rho}=5.0$ is obtained from the decay of $\rho \to e^{+}e^{-}$ and $g_{\rho}=6.0$ from the 
strong decay of $\rho \to \pi^{+}\pi^{-}$ 
\cite{Volkov:1986zb, Gounaris:1968mw, Ebert:1982pk, Klingl:1996by, Li:1995tv, Benayoun:2007cu}. 
In the present work, we have succeeded to show that the decays $\rho \to e^{+}e^{-}$ and 
$\rho \to \pi^{+}\pi^{-}$ can be described using only one value $g_{\rho}=5.0$. This was achieved 
by taking into account the interaction of pions in the final state in the $\rho \to \pi^{+}\pi^{-}$ decay. 
That also made it possible to describe the decay 
$\tau^{-} \to \pi^{-}\pi^{0} \nu_{\tau}$ and the process $e^{+}e^{-} \to \pi^{+}\pi^{-}$ in satisfactory 
agreement with the experimental data.

The last two processes were considered in the NJL model in our previous work~\cite{Volkov:2020uld}. 
In that paper interactions in the final state also taken into account, however, for the vector coupling 
constant, the value $g_{\rho}=6.14$ was used and the parameter $\Lambda_M =740$ MeV was 
fixed according to the experiment $e^{+}e^{-} \to \pi^{+}\pi^{-}$. In the present work,
we fix the value of the parameter $\Lambda_M =880$~MeV using the experimental value for the decay width 
$\rho \to \pi^{+}\pi^{-}$. 

In this case, quite satisfactory predictions are obtained for the decay width 
$\tau^{-} \to \pi^{-}\pi^{0} \nu_{\tau}$ and the cross section of the process  $e^{+}e^{-} \to \pi^{+}\pi^{-}$.

Thus, we have shown that in the considered processes, taking into account the interaction of pions 
in the final state plays a very important role. This effect can be described by a meson loop, which 
takes into account the exchange of outgoing pions by the $\rho$ meson in the P wave state. Note 
that the importance of taking into account the interaction of pions in the final state is due to the 
relative smallness of the energy of the emitted pions. Satisfactory agreement with experimental 
data for all considered processes confirms the applicability of the vector current conservation 
hypothesis.


\begin{thebibliography}{32}

\bibitem{Volkov:1986zb} 
  M.~K.~Volkov,
  Low-energy Meson Physics in the Quark Model of Superconductivity Type. (In Russian),
  Sov.\ J.\ Part.\ Nucl.\  {\bf 17}, 186 (1986)
  [Fiz.\ Elem.\ Chast.\ Atom.\ Yadra {\bf 17}, 433 (1986)].
    
\bibitem{Ebert:1985kz} 
  D.~Ebert and H.~Reinhardt,
  Effective Chiral Hadron Lagrangian with Anomalies and Skyrme Terms from Quark Flavor Dynamics,
  Nucl.\ Phys.\ B {\bf 271}, 188 (1986).
  
\bibitem{Vogl:1991qt} 
  U.~Vogl and W.~Weise,
  The Nambu and Jona Lasinio model: Its implications for hadrons and nuclei,
  Prog.\ Part.\ Nucl.\ Phys.\  {\bf 27}, 195 (1991).
  
\bibitem{Klevansky:1992qe} 
  S.~P.~Klevansky,
  The Nambu-Jona-Lasinio model of quantum chromodynamics,
  Rev.\ Mod.\ Phys.\  {\bf 64}, 649 (1992).
  
\bibitem{Hatsuda:1994pi}
  T.~Hatsuda and T.~Kunihiro,
  Phys.\ Rept.\  {\bf 247}, 221 (1994).

\bibitem{Volkov:2020uld}
  M.~K.~Volkov, A.~B.~Arbuzov and A.~A.~Pivovarov,
  Pis'ma ZhETF {\bf 112}, 493 (2020) 
  [arXiv:2010.04581 [hep-ph]].

\bibitem{Tanabashi:2018oca} 
  M.~Tanabashi {\it et al.} [Particle Data Group],
  Review of Particle Physics,
  Phys.\ Rev.\ D {\bf 98}, no. 3, 030001 (2018).

\bibitem{Vainshtein:1970zm} 
  A.~I.~Vainshtein and V.~I.~Zakharov,
  Sov.\ Phys.\ Usp.\  {\bf 13}, 73 (1970)
  [Usp.\ Fiz.\ Nauk {\bf 100}, 225 (1970)].
  
\bibitem{Achasov:2005rg} 
  M.~N.~Achasov {\it et al.},
  J.\ Exp.\ Theor.\ Phys.\  {\bf 101}, no. 6, 1053 (2005)
  [Zh.\ Eksp.\ Teor.\ Fiz.\  {\bf 128}, no. 6, 1201 (2005)]
  [hep-ex/0506076].
  
\bibitem{Ablikim:2015orh} 
  M.~Ablikim {\it et al.} [BESIII Collaboration],
  Phys.\ Lett.\ B {\bf 753}, 629 (2016)
  [arXiv:1507.08188 [hep-ex]].
  
\bibitem{Gounaris:1968mw} 
  G.~J.~Gounaris and J.~J.~Sakurai,
  Phys.\ Rev.\ Lett.\  {\bf 21}, 244 (1968).

\bibitem{Ebert:1982pk}
  D.~Ebert and M.~K.~Volkov,
  Z.\ Phys.\ C {\bf 16}, 205 (1983).
    
\bibitem{Li:1995tv} 
  B.~A.~Li,
  Phys.\ Rev.\ D {\bf 52}, 5184 (1995)
  [hep-ph/9505235].
  
\bibitem{Klingl:1996by}
  F.~Klingl, N.~Kaiser and W.~Weise,
  Z.\ Phys.\ A {\bf 356}, 193 (1996)
  [hep-ph/9607431].
  
\bibitem{Benayoun:2007cu} 
  M.~Benayoun, P.~David, L.~DelBuono, O.~Leitner and H.~B.~O'Connell,
  Eur.\ Phys.\ J.\ C {\bf 55}, 199 (2008)
  [arXiv:0711.4482 [hep-ph]].

 \end{thebibliography}
\end{document}